\documentclass[%
notitlepage, 10pt, twocolumn,
superscriptaddress,
amsmath,amssymb,
aps,prx,
floatfix,
]{revtex4-2}

\usepackage[dvipsnames]{xcolor}
\usepackage{hyperref}
\hypersetup{
  breaklinks=true,
  colorlinks=true,
  allcolors=BlueViolet,
}
\usepackage{url}
\usepackage{float}

\usepackage{physics}
\usepackage{amsthm}
\usepackage{bm}% bold math
\usepackage{graphicx}% Include figure files
\graphicspath{{./figures/}} % set path for all figures
\usepackage{dcolumn}% Align table columns on decimal point
\usepackage[linesnumbered,boxed]{algorithm2e}
\usepackage{mathtools}

\usepackage{braket}
\usepackage{nccmath}  % for \mfrac

\newcommand{\p}[1]{\left(#1\right)}
\renewcommand{\sp}[1]{\left[#1\right]} % square parenthesis

\newcommand{\D}{\mathcal{D}}
\renewcommand{\L}{\mathcal{L}}
\renewcommand{\O}{\mathcal{O}}

\renewcommand{\o}{\text{o}}

\def\oket#1{\mathinner{|{#1})}}

\usepackage{dsfont}
\newcommand{\1}{\mathds{1}}

\newcommand{\black}[1]{\textcolor{black}{#1}}

\begin{document}

\title{Achieving the Heisenberg limit with Dicke states in noisy quantum metrology}

\author{Zain H. Saleem}
\email{zsaleem@anl.gov}
\affiliation{Mathematics and Computer Science Division, 
Argonne National Laboratory, 9700 S Cass Ave, Lemont IL 60439}

\author{Michael Perlin}
\email{mika.perlin@gmail.com}
\affiliation{Infleqtion, Inc., Chicago, IL 60615, USA}

\author{Anil Shaji}
\email{shaji@iisertvm.ac.in}
\affiliation{School of Physics, IISER Thiruvananthapuram, 
Kerala, India 695551}

\author{Stephen K. Gray}
\email{gray@anl.gov}
\affiliation{Center for Nanoscale Materials,
Argonne National Laboratory, Lemont, Illinois 60439, USA}

\begin{abstract}
Going beyond the standard quantum limit in noisy quantum metrology is an important and challenging task. Here we show how Dicke states can be used to surpass the standard quantum limit and achieve the Heisenberg limit in open quantum systems. The system we study has qubits symmetrically coupled to a resonator and our objective is to estimate the coupling between the qubits and the resonator. The time-dependent quantum Fisher information with respect to the coupling is studied for this open quantum system where the same decay rates are assumed on all qubits. We show that when the system is initialized to a Dicke state with an optimal excitation number one can go beyond the standard quantum limit and achieve the Heisenberg limit even for finite values of the decays on the qubit and the resonator, particularly when the qubits and resonator are strongly coupled.
\textcolor{black}{We compare our results against the highly entangled GHZ state and a completely separable state and show that the GHZ state performs quite poorly whereas under certain noise conditions the separable state is able to go beyond the standard quantum limit due to subsequent interactions with a resonator.
%due to the entanglement introduced by the dynamics.
}
\end{abstract}

\maketitle

\section{Introduction} %--
%Quantum metrology concerns the task of estimating a parameter, or several parameters, characterizing the Hamiltonian of a quantum system 
A quantum probe with $N$ elementary sub-units is the focus of discussions on quantum-limited single parameter estimation \cite{giovannetti2011advances, toth2014quantum, nawrocki2015introduction, polino2020photonic}. The probe is chosen such that the parameter to be estimated appears in its Hamiltonian. A suitable initial state of the probe evolves for a fixed time  in a manner that depends on the parameter value. The value can be then inferred from a read-out of the final state of the probe.
%This task is performed by preparing a suitable initial state of the system, allowing it to evolve for a specified time, performing a suitable measurement, and inferring the value of the parameter(s) from the measurement outcome. 
Quantum mechanics places a fundamental limit on measurement precision, called the Heisenberg limit (HL), which constrains how the precision of parameter estimation improves as the number of probe units increases. According to the HL, scaling of the precision with the number of elementary sub-units cannot scale better than $1/N$. For a noiseless system, HL scaling can be saturated using entangled states of the probe 
\cite{giovannetti2004quantum}. 
In practice, though, environmental decoherence typically degrades 
metrologically useful entanglement; instead of HL, precision 
scales like $1/\sqrt{N}$, called the standard quantum limit (SQL), 
which can be achieved by using the $N$ elementary sub-units independently. \textcolor{black}{
The burgeoning area of noisy quantum metrology seeks to characterize
and mitigate the detrimental effects of noise
\cite{pezze2008mach, plenio2016sensing, Rafal2017, Zhou2018, dur_improved_2014,huang_imaging_2022, kessler_quantum_2014, layden_ancilla-free_2019, shettell_practical_2021, shettell2020graph, matsuzaki2011magnetic, chin2012quantum, altherr_quantum_2021, Huelga1997, Chaves2013, Frowis2014, Altenburg2016, saleem2022optimal}.
For example, to surpass the SQL various clever strategies have been devised,
such as
squeezing the vacuum \cite{pezze2008mach},
monitoring the environment
during the measurements \cite{plenio2016sensing}, quantum error correction
\cite{Rafal2017,Zhou2018,dur_improved_2014,huang_imaging_2022,
kessler_quantum_2014,layden_ancilla-free_2019,shettell_practical_2021},
the use of graph states \cite{shettell2020graph},
and taking advantage of non-Markovian
dynamics \cite{matsuzaki2011magnetic,chin2012quantum,altherr_quantum_2021}
of the probe.
} % end of textcolor{black}{ above

The quantum Cramer-Rao bound \cite{cramer1946contribution}  dictates that the precision in the measurement of a parameter in a  quantum experiment is bounded by the inverse of the quantum Fisher information (QFI).
\textcolor{black}{The QFI is a time-dependent quantity which, in open quantum systems, first grows as the target signal gets imprinted into the system, before decreasing as decoherence randomizes the system and degrades the signature of the target signal.
The best precision in the estimate of a parameter can, in such cases, be achieved by performing measurements near the times when the QFI is the largest \cite{Altenburg2016,saleem2022optimal}.
}

%\textcolor{black}{
%For closed quantum systems the QFI is a  monotonically increasing function of time because the statistical distance between states that start out infinitesimally close to each other keeps growing.
%However when there is decoherence the statistical distance cannot keep on growing in time and therefore has a peak \cite{Altenburg2016,saleem2022optimal}.
%The time dependence of the QFI was studied in Refs.~\cite{Altenburg2016,saleem2022optimal} and, as expected, the QFI often has a maximum as a function of time in open quantum systems due to the competition between the parameter-dependent unitary evolution and the noise.
%}

\textcolor{black}{In this paper, we introduce a strategy for surpassing the SQL in a noisy quantum system by making use of an optimal measurement time. We numerically explore the advantages of using the optimum measurement time when the quantum probe is initialized in  Dicke states \cite{Dicke1954,shammah2018open} with appropriately chosen excitation numbers. We also compare the measurement precision and its scaling obtained using Dicke states with that of highly entangled states like the GHZ-state on one hand and simple X-polarized product states on the other, and find that in certain noise regimes, when optimizing over measurement time, even the product state can yield better than SQL scaling.}

We investigate a qubit-cavity model in contact with an environment where both the qubits and the cavity are allowed to decay under the influence of noise from the environment. The parameter that is estimated is the coupling constant of the qubits to the cavity. The qubits constitute the quantum probe and we focus on studying the scaling of the uncertainty in its estimate with the number of qubits, $N$. We calculate the scaling of the time-optimized QFI with $N$ and see for what choices of the initial states and decay rates of the qubits and the resonator we are able to surpass the SQL. The Hamiltonian of our system is permutationally symmetric, and this symmetry allows us to use a basis that reduces the dimensions of our problem and thus allows us simulate open quantum systems with relatively large values of $N$ that would otherwise be computationally intractable. Dicke states are examples of permutationally symmetric states and we will show that when the excitation number of the Dicke states is chosen appropriately and the measurements are performed in an optimal time region, we can surpass the standard quantum limit even in the presence of dissipation.
%We attribute such results to a competition between  degree of entanglement in the initially prepared probe state with its resilience to noise \textcolor{black}{coupled with the optimal measurement time}.
%\textcolor{black}{We attribute surpassing the SQL in the presence of noise to the optimization over measurement time, which may differ for different system sizes and decoherence rates.}

\textcolor{black}{It is important to note that there are ``no-go’’
theorems and also particular scenarios  for which it has been
established that open systems described by the Lindblad master equation
cannot exhibit scaling beyond the standard quantum limit.
%For example, Refs.~\cite{Rafal2017, Zhou2018} establish that for
%fixed total probing times with Lindblad dynamics and
%Hamiltonians such as the one we explore, that the standard
%quantum limit cannot be overcome, but error correction schemes
%involving ancilla qubits can ameliorate this situation.
%Our focus is on measurements at times when the quantum Fisher
%information attains maximum values and, furthermore,
%these times are not fixed but vary with the number of component qubits.
%In the following we investigate the case where the measurement is
%always performed at the optimal time for all the states and
%parameters considered. We also consider a more generic noise model
%described by a GKSL type master equation. Choosing the optimal time
%allows us to circumvent these no-go theorems and with losses
%on both qubits as well as the cavity, it is still possible to go
%beyond SQL scaling.
In particular, Refs. \cite{Rafal2017, Zhou2018}  establish for systems such as we explore that an upper limit on the QFI scales linearly with the total probing time in a sequential scheme, which implies that the best one could hope to achieve is the SQL.  If one were to assume that the corresponding total probing time in the parallel, $N$-qubit case is proportionate to $N$ times some \textit{fixed}  (smaller) probing time, then one would also expect no better than SQL scaling with $N$.  Our focus, however, is the case where measurements are performed at optimal times when the QFI is maximum, and these times vary with $N$.  That these probing times are not fixed with $N$ allows us to circumvent the no-go theorems.  Of course,  a negative aspect of our approach is that it is more complicated than schemes with fixed probing times but we have suggested an experimental protocol that can offer some efficiencies \cite{saleem2022optimal}.
}  % end textcolor

The significance of our results and our main motivation lies in the fact that Dicke states with modest excitation numbers can be realised in the laboratory \cite{zou2018beating, chen_-chip_2023,chiuri_hyperentangled_2010,haas_entangled_2014,haffner_scalable_2005,hume_preparation_2009,noguchi_generation_2012,prevedel_experimental_2009,toyoda_generation_2011,wieczorek_experimental_2009,yu_observation_2019} and exploring the interplay between noise and metrological precision can be investigated experimentally. \textcolor{black}{Indeed, there are a few proposals for using Dicke states and closely related states in specific quantum metrology schemes, including noisy quantum
metrology, as well~\cite{Altenburg2016,paulisch_quantum_2019,zhang_quantum_2014,Frowis2014}.  See
also the review articles  Refs. \cite{pezze2018quantum} and \cite{polino2020}
for additional discussion of Dicke states in quantum metrology.} In addition there are multiple theoretical proposals for producing such states in the laboratory \cite{aktar_divide-and-conquer_2022, chakraborty_efficient_2014, ivanov_creation_2013, kasture_scalable_2018, lamata_deterministic_2013, mu_dicke_2020, raghavan_generation_2001, stojanovic_dicke-state_2023, wang_preparing_2021, wei_deterministic_2015, wu_generation_2017, zhao_efficient_2011}, giving further relevance to our work.

\subsection{Quantum Fisher information} %--

Consider the problem of estimating a (scalar) parameter $\theta$ in a quantum experiment and denote the (unbiased) estimator of this parameter by $\hat{\theta}$.
The parameter dependent evolution of the probe is given by $H_{\rm probe} = \theta h$, where $h$ is an operator on the probe state.
The variance in the estimate of the parameter, $\delta \hat \theta ^2$,  is upper-bounded by the quantum Cramér-Rao bound \cite{holevo2011probabilistic}, 
\begin{align}
  \delta \hat \theta ^2 \geq \frac{1}{M F(\theta)}  
\end{align}
where $M$ stands for the total number of application of the quantum probe and $F(\theta)$ is the QFI, which quantifies the responsiveness of the quantum state of the probe to changes in the measured parameter $\theta$. The QFI of a mixed state $\rho = \sum_a \lambda_a \op{\lambda_a}$ with respect to the parameter $\theta$ is \cite{liu2019quantum}
\begin{align}
    \label{fisher1}
F(\theta) = 2 \sum_{a,b}
\frac{\abs{\braket{\lambda_a|\partial_\theta\rho|\lambda_b}}^2}{\lambda_{a}+\lambda_{b}},
\end{align}
where the sum is taken over values of $a,b$ for which $\lambda_a + \lambda_b\ne0$.
%where the sum is implicitly taken over values of $a,b$ with nonzero eigenvalues $\lambda_a,\lambda_b$. Since $\rho$ is, in general, time-dependent,  the QFI also has time dependence. For closed quantum systems with no losses or decays present in the system the QFI is a monotonically increasing function with time.
For closed evolution of the probe, the QFI has a quadratic time dependence while for open evolution an optimal measurement time may be present that the parameter estimation protocol would take advantage of \cite{saleem2022optimal}.
%quantum parameter estimation can benefit from performing measurements at times where the QFI is maximum. However, for open quantum systems the QFI has a more complex behavior with time and can exhibit a maximal value in time. 
We  study the scaling of the peak value of QFI under open evolution as $N$ gets larger. 

\section{Model system}%--
The qubit-cavity model that we study here is described by the Hamiltonian ($\hbar$ = 1)
\begin{align}
\label{Hamil}
H = \omega_q S_z + \omega_c a^{\dagger} a + g(a^{\dagger} S_- + a S_+),
\end{align}
where $S_z = \frac12 \sum_{j=1}^N \sigma_z^{(j)}$, and $S_\pm = \sum_{j=1}^N \sigma_\pm^{(j)}$ are collective spin operators expressed in terms of the \black{single-qubit Pauli matrices $\sigma_z = \op{\uparrow} - \op{\downarrow}$, $\sigma_+ = \op{\uparrow}{\downarrow}$, and $\sigma_- = \sigma_+^\dag$}; $a^\dag$ and $a$ are bosonic raising and lowering operators for the cavity; $\omega_q$ is the qubit excitation energy; $\omega_c$ is the resonator frequency; and $g$ is the qubit-cavity coupling strength. Our focus will be on the QFI with  respect to measurement of the coupling constant, $g$, i.e., $\theta = g$ in Eq.~\eqref{fisher1}. Since we are only concerned with estimating the coupling, we work on resonance and in a rotating frame, or equivalently set $\omega_q=\omega_c=0$.
%will set the detuning $\Delta$ and the resonator frequency $\omega_c$ to zero by tuning $\omega_c=\omega_q$ and going into the appropriate rotating frame. 
The state of the quantum probe undergoes open evolution described by the Gorini-Kosskowski-Sudarshan-Lindblad (GKSL) equation \cite{gorini_completely_1976, lindblad_generators_1976, chruscinski_brief_2017, manzano2020short},
\begin{align}
\dot{\rho} = -i [ H, \rho ] + \kappa \D\left[a\right](\rho) + \gamma \sum_{j=1}^N \D\left[\sigma_-^{(j)}\right](\rho),
\label{Lindblad2}
\end{align}
where $\kappa$ and $\gamma$ are, respectively, resonator and qubit decay rates, and the dissipator
\begin{align}
  \D[\O](\rho) = \O \rho \O^\dag - \frac12 (\O^\dag \O \rho + \rho \O^\dag \O).
\end{align}
%Most of our simulations were performed with the Quantum Toolbox in Python (QuTiP) \cite{qutip1, qutip2}.
%  2/21/2024, SKG: for some reason, this important statement about how things scale
% was deleted.  I noticed this when looking at reviewer B's final
% remarks that asked about this very statement which was in
% fact no longer there and yet we responded!!??
With the system as described, the dynamics can be expressed in
terms of the scaled, unitless variables $gt$, $\kappa/g$, $\gamma/g$ and
\textcolor{black}{
the QFI (with respect to parameter $g$) can be represented
as the dimensionless product, $F(g) \times g^2$. See Appendix \ref{sec:numerics} for additional details on scalings. } % end textcolor

In principle, if the cavity excitation number is bounded above by the qubit number $N$, the Hilbert space of the combined qubit-cavity system has dimension $(N+1) 2^N$.
Operators on this Hilbert space (such as $\rho$), in turn, have $O(N^2 4^N)$ degrees of freedom, which severely limits computational capability even for moderately large $N$.
However, spin-permutation symmetry\black{---namely, the invariance of $\rho$ under arbitrary permutations of its spins---reduces the degrees of freedom for a mixed state of $N$ spins to a modest $O(N^3)$ \cite{chase2008collective, baragiola2010collective, moroder2012permutationally, novo2013genuine, xu2013simulating, bolanos2015algebraic, shammah2017superradiance, zhang2018montecarlo, shammah2018open}, allowing us to perform calculations in a symmetric basis for the combined spin-boson system that grows as $O(N^5)$}.
Though individual terms on the right hand side of Eq.~\eqref{Lindblad2} may break permutational symmetry, the sum of these terms does not.
If $\rho$ is initially invariant under all permutations of spins, therefore, this symmetry is preserved at all times.
This symmetry allows us to expand $\rho$ in a basis of matrix elements of the form $\op{J,n}{J,m}\otimes \op{\ell}{k}$, where \textcolor{black}{$J\in\{\frac{N}{2},\frac{N}{2}-1,\cdots,\frac{N\!\!\mod2}{2}\}$ is a total spin length}, $n,m\in\{J,J-1,\cdots,-J\}$ are spin projections onto the spin quantization axis, and $\ell,k\in\{0,1,\cdots,N\}$ are cavity occupation numbers.
Though the spin quantum numbers $J$ and $m$ are generally insufficient to uniquely determine the state of $N$ spins when $J<N/2$, the values of additional quantum numbers are fixed by the requirement of spin-permutation symmetry \cite{shammah2018open}.
\black{
We provide additional details about our numerical simulations (benchmarked in small-scale systems against QuTiP \cite{johansson2012qutip, JOHANSSON20131234}) in Appendix \ref{sec:numerics}, and make our codes publicly available at Ref.~\cite{perlin2023spinboson}.
}

%The permutational symmetry of our basis restricts the type of initial state we can use in our computation. 
%As we have chosen our coupled basis states to be such that the dimensions of density matrix have been reduced, we need to choose the initial states that are also permutationally symmetric. 
%We consider three types of permutationally symmetric initial states for the qubits. These are the the n-excitation dicke states, 
%In addition to the Dicke-$n$ initial states, for comparison we also consider
%GHZ initial qubit states and  (un-entangled) maximal superposition qubit states (or polarized-$X$ states), both of which are permutationally  symmetric and defined below. 

\begin{figure}
  \centering
  \includegraphics{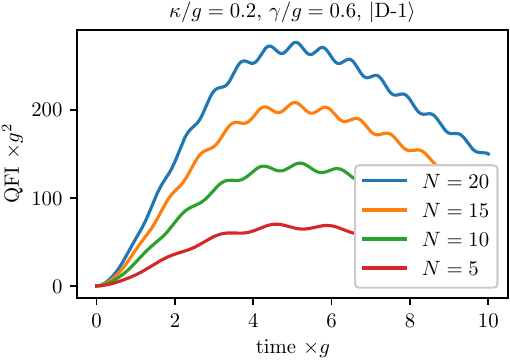}
  \caption{Time-dependent QFI for different qubit numbers, $N$,  for an initial Dicke-1 state with relatively small qubit ($\gamma$) and resonator ($\kappa$) decay rates. The peak structures increase with $N$ for this case.}
  \label{time}
\end{figure}

Our main focus is on the case where the quantum probe is initialized in Dicke states of the qubits with resonator in its ground state,
\begin{align}
  \ket{\text{D-}n} = \ket{\phi_n} \otimes \ket{0},
  &&
  \ket{\phi_n} \propto S_+^n \ket{0}^{\otimes N}.
\end{align}
Here $\ket{\phi_n}$ is a normalized Dicke state of $N$ qubits, which can be equivalently defined as a uniform superposition of all $N$-qubit states with exactly $n$ excitations, e.g.,
\begin{align}
  \ket{\phi_1} = \frac1{\sqrt{N}}
  \big(\ket{100\cdots} + \ket{010\cdots} + \ket{001\cdots}\big).
\end{align}
$|\phi_1\rangle$ is also known as the $N$-qubit W state \cite{Dur2000}.
For ease of language, we generally refer to $\ket{\text{D-}n}$ as the Dicke-$n$ state of the qubit-resonator system.

For comparison with the $\ket{\text{D-}n}$ state we also consider the $X$-polarized state
\begin{align}
  \ket{\text{X}}
  = \black{\p{\frac{\ket{0}+\ket{1}}{\sqrt{2}}}^{\otimes N} \otimes \ket{0}
  = \sum_m c_{Nm} \Ket{\mfrac{N}{2},m} \otimes \ket{0}},
\end{align}
\black{where $c_{Nm}=\sqrt{{N\choose N+m}/2^N}$,} and the GHZ state
\begin{align}
  \ket{\text{GHZ}}
  &= \frac1{\sqrt{2}} \p{\ket{0}^{\otimes N} + \ket{1}^{\otimes N}} \otimes \ket{0} \nonumber \\
  &\black{= \frac1{\sqrt{2}}\Big(\Ket{\mfrac{N}{2},\mfrac{N}{2}} + \Ket{\mfrac{N}{2},-\mfrac{N}{2}}\Big) \otimes \ket{0}},
\end{align}
\black{where we provide these states in the collective $\ket{J,m}$ spin basis, and both states are} defined with the resonator in its ground state.

\begin{figure}
    \resizebox{8.5cm}{6 cm}{\includegraphics{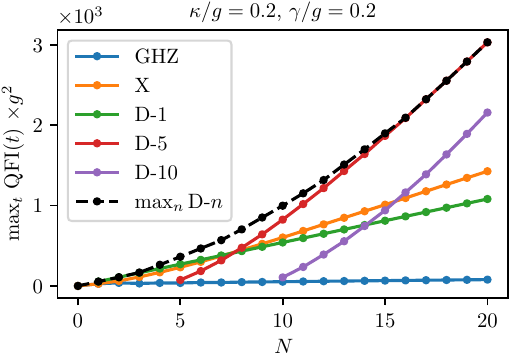}}
    \resizebox{8.5cm}{6 cm}{\includegraphics{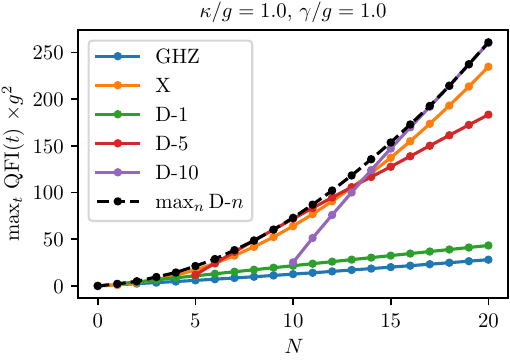}}
    \caption{Scaling of the maximum QFI with respect to time for fixed qubit and resonator decay rates, and different initial states. The upper and lower panels  correspond to small and large values of the  decay rates, respectively. The upper panel with $\kappa$, $\gamma \ll g$, corresponds to strong coupling between the qubits and the cavity while the lower panel with $\kappa$, $\gamma \sim g$ corresponds to moderate coupling.
    \textcolor{black}{Dashed black line shows, for each qubit number $N$, the time-optimized QFI maximized over choice of initial Dicke state.}
    }
    \label{scaling}
\end{figure}

%The QFI with $N$ elementary sub-units or qubits can exhibit multiple peaks at different times and as one changes $N$ different peaks compete for the global maximum. 
\section{Time-dependence of the QFI for $N$ qubits}%--

In Fig.~\ref{time} we show a plot of the time-dependent QFI for the Dicke-1 initial state for different values of $N$, keeping the decay constants on all the qubits and the resonators fixed.  One sees that on average the QFI rises and then falls in time, with some oscillatory substructure superimposed.  The region of largest QFI values and also the associated overall maximum QFI value increase in magnitude with increasing $N$.   This behavior is not limited to the Dicke-1 state and happens for all the other Dicke-$n$ states and for the GHZ and $X$-polarized initial states as well \textcolor{black}{(see Appendix \ref{sec:timedependence} for examples)}.

%This gives rise to the appearance of various cusps in the plot of the maximum value of the QFI vs $N$ and fitting those curves to a smooth line is not perfect. 
% SKG: the new Fig. 2 has no evident cusps and so I commented out the remark here
% and also above about cusps and imperfect fits.
In Fig.~\ref{scaling} we provide plots of how the maximum QFI value scales with $N$. Note that we use $N$ to quantify the resources that go into the measurement of $g$ even though, strictly speaking, we have $N$ qubits plus a resonator whose Hilbert space dimension grows as $N+1$.
%It can be seen that the lines are not perfectly smooth because of the shift of the peak corresponding to the global maxima as we change $N$ as we have noted earlier. We will however ignore these cusps and fit a smooth curve to them to get an ``approximate scaling'' of the QFI with respect to $N$. The cusps in the curves are not very drastic and so we can rely on this approximate scaling. 
It is interesting to note in Fig.~\ref{scaling} that the GHZ state, despite being maximally entangled by some measures, scales very poorly for both small and large values of decay rates. This is due to the fragility of the GHZ state to decoherence. For example, measuring just one qubit collapses the  GHZ state into a  separable state, whereas this is not the case for the W state or, indeed, most of the Dicke states.  
\textcolor{black}{
Of course this possible limitation of the GHZ state for quantum sensing has been noted many times, which has been motivation of exploration of
other entangled states, not just Dicke states, such as
graph states \cite{shettell2020graph} and also motivation for the development of schemes
to mitigate the effects of noise, e.g. Refs. \cite{Rafal2017, Zhou2018,Chaves2013,Altenburg2016}.
} % end textcolor
% SKG: the following sentence is not always true in Fig. 2 (see large N, lower
% panel) and is also a little out of place. So I commented it out:
%The Dicke states with larger excitation numbers are more resilient to noise in terms of scaling but the magnitude of the QFI for these states is lower. 
For large values of the decay constants (lower panel of Fig.~\ref{scaling}) the initially unentangled $X$-state scales in a manner almost comparable to the large excitation Dicke state.
\textcolor{black}{It is worth noting that the SQL, which applies to statistically independent probes, does not truly apply to this initial state because the spins interact with a shared resonator.}
We note, however, that the $X$-state is only competitive with the best Dicke states in the limit of relatively high dissipation and low QFI magnitudes, i.e. in a limit less well suited to achieving high-precision sensing. For a metrology performance comparison between separable states like the $X$-state and maximally entangled states see Ref.~\cite{beau2017nonlinear}.

\begin{figure}
    \resizebox{8.5cm}{6cm}{\includegraphics{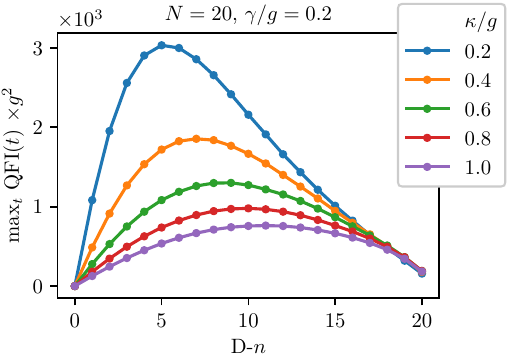}}
    \caption{Variation of the time-optimized QFI with Dicke excitation number for $N=20$ qubits and varying resonator decay rate, $\kappa$, for fixed qubit decay rate, $\gamma$.}
    \label{MaxQFIvsn}
\end{figure}

\subsection{Optimal excitation number of Dicke states}%--

Fig.~\ref{scaling} shows that, depending on the number of
qubits, $N$, there can be optimal Dicke state excitation numbers. In the case of low dissipation (upper panel of Fig.~\ref{scaling}) for example, Dicke-1 has the larger QFI for  $N=5$--10,  whereas  Dicke-5 has the higher QFI for for $N > 10$.
In Fig.~\ref{MaxQFIvsn} we plot the maximum value of the QFI against different Dicke excitation numbers for $N = 20$ and see that there is indeed an optimal value of the excitation number for which the maximum value of the QFI is the largest for each $\kappa$ considered. 
By the geometric entanglement measure, for fixed $N$, it turns out that the most entangled Dicke state is equal to or close to Dicke-$N/2$ \cite{Martin2010}. For the $N$ = 20 case of Fig.~\ref{MaxQFIvsn}, this would correspond to Dicke-10.  The results of Fig.~\ref{MaxQFIvsn}, however, show optimum Dicke states ranging from Dicke-5 at the lowest $\kappa$ considered to Dicke-11 at the highest $\kappa$ considered, again consistent with entanglement alone not being the best gauge of sensing quality and that a competition between degree of entanglement and resilience to noise leads to the optimum initial probe state.

\textcolor{black}{Note that by focusing on measurements at times near the maximum of the QFI$(t)$ we are not explicitly considering the total measurement time as a resource to be minimized. If the times associated with individual shots are constant or not a significant resource, e.g. if state preparations, measurements or other fixed experimental times dominate over actual system evolution times, then this is valid. Indeed, in previous work some of us showed that an experimental protocol can be developed using our optimum time idea that significantly reduces the total number of shots required to achieve a given precision in a parameter \cite{saleem2022optimal}. However, if the total measurement time associated with all the shots is considered a resource, and if for simplicity one neglects state preparation and other experimental factors, then one must consider QFI$(t)/t$ as the figure of merit to be maximized with respect to $t$ \cite{Chaves2013} (See also Ref.~\cite{Huelga1997} that minimized an associated uncertainty with measurement time in mind). Another scenario in which the total time is relevant is when the measured parameter is also time dependent and there is a small window of time within which the value of the parameter has to be estimated. However in our case, the coupling $g$ that we are estimating is constant in time and such considerations do not apply. For completeness, we have  also carried out a number of calculations using maxima associated with QFI$(t)/t$  (see Appendix \ref{sec:QFIovertime}) and have found very similar results to those described here in the main text.
}

\begin{figure}[!htb]
  \resizebox{6.5cm}{5.2cm}{\includegraphics{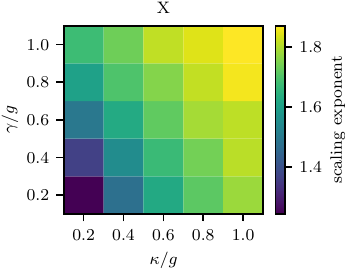}}
  \resizebox{6.5cm}{5.2cm}{\includegraphics{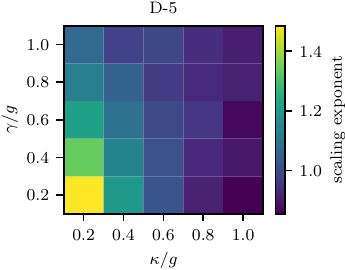}}
  \resizebox{6.5cm}{5.2cm}{\includegraphics{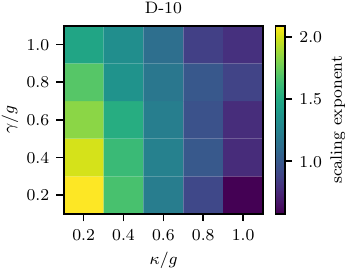}}
 \caption{The scaling exponent $b$ obtained by fitting the time-maximized QFI to the qubit number $N$ as $\max_t\text{QFI}(t,N) = aN^b+c$ for different resonator and qubit decay rates $\kappa$ and $\gamma$ and three initial states. We see that when the qubits are strongly coupled to the cavity (relative to the losses), the Dicke states perform very well in terms of the scaling of the QFI with $N$ while the initially un-entangled X-state performs better in the moderate coupling regime.}
  \label{mapplots}
\end{figure}

Finally, we consider if certain ranges of the decay constants associated with the qubits and the resonator can lead to scaling that surpasses the standard quantum limit or SQL. To accomplish  this, we fit $y(N) = a N^b +c$  to the maximum QFI vs $N$ plots for a large range of  decay constant values and extract the scaling exponent, $b$. The results of our numerical simulations are summarized in Fig.~\ref{mapplots} where we plot the exponent $b$ for different values of the decay constants for three different initial states: the $X$-state, the Dicke-5, and Dicke-10. In these plots $b = 1$ corresponds to the SQL and $b = 2$ corresponds to the Heisenberg limit.  Regarding the initially  unentangled $X$-state, scaling closer to the SQL limit occurs for low decay constant values, as might  be expected.  A transition to much better than SQL scaling occurs for higher decay constants, possibly due to dissipation induced entanglement. However, as previously noted, the actual magnitudes of the QFI can be orders of magnitude smaller than what is achievable with Dicke states.
For the  Dicke states, it can be clearly seen that for certain, low but finite values of the decay rates the Dicke states are able to go above the SQL and even achieve the Heisenberg limit.

\textcolor{black}{In some previous work on noisy quantum metrology, 
\cite{zhang_quantum_2014,kurdzialek_using_2023,
liu_optimal_2023,nichols_practical_2016,perarnau-llobet_weakly_2021,
wang_quantum-enhanced_2022,zhang_demonstrating_2019,zhou_optimal_2023,
shettell2020graph} the importance of finding probe states that are 
resilient to noise rather than just highly entangled was pointed out. 
but the use of the optimal time for sensing was not considered.
The use of optimum time ideas can be found in 
Refs. \cite{Chaves2013,Frowis2014, Altenburg2016}, but for much simpler model systems
and different noise models than us. Reference \cite{Chaves2013} was able to show
Heisenberg limit scaling with GHZ states could be achieved for their system and noise model. References \cite{Frowis2014, Altenburg2016}
also made use of Dicke states but neither of these references was able to go beyond the SQL by making use of optimum time. Reference
\cite{Altenburg2016} was able to achieve HL in a steady state limit where the collective phase noise considered in the model does not change the state and with
auxiliary qubits via differential interferometry.
Furthermore, earlier work suggests that separable probe states 
can never go beyond the SQL scaling. However, in our work we have 
demonstrated 
%for the first time 
that for some models the subsequent dynamics allows the separable probe states to go beyond the SQL scaling and even achieve the Heisenberg limited scaling when the strategy of optimising the measurement time is also employed. 
In Ref. \cite{Altenburg2016} it was pointed out that in the presence of correlated phases noise,
X-states can perform better than entangled ones. However, the scaling
of the measurement uncertainty with $N$ when the optimal time is
always used was not considered.}  

It should be noted that the Dicke state HL limit results in Fig.~\ref{mapplots} correspond to parameters in the strong coupling limit of $g \gg \kappa,~ \gamma$ which can sometimes be hard to achieve in practice. Interestingly the $X$-state results show near HL limit scaling in a somewhat weaker or moderate coupling regime of $g$ $\approx$  $\kappa,~ \gamma$.

\section{Conclusions and Future Directions}%--

\textcolor{black}{Many studies in noisy quantum metrology focus on uncoupled or weakly coupled
systems of qubits.  In contrast we study here systems of qubits coupled to
a bosonic resonator (which in turn mediates qubit couplings). The dynamics
of such systems, after preparation of some initial state, can be nontrivial
and put to metrological advantage. We map out the behavior of such systems over a range of loss parameters for the qubits
and the resonator, identifying optimal initial states and parameter ranges
for going beyond the SQL. Among our interesting results include the use
of Dicke states optimized with respect to excitation number, and even the
observation that initially separable states can yield scaling beyond the SQL
under certain conditions.  
In coming to these conclusions we have made use of previously developed
concepts in noisy quantum metrology, including the concept of optimal measurement
times \cite{Huelga1997,Chaves2013,Frowis2014,Altenburg2016,saleem2022optimal} 
and the general notion that some entangled states such as Dicke
states can be more robust to noise \cite{Frowis2014,Altenburg2016,
shettell2020graph}.}

\black{For future work we will like to extend our study to include other noise models such as dephasing. Implementing our proposal to surpass the standard quantum limit in a laboratory can be a fruitful direction of research as well. The ability to produce Dicke states in the laboratory \cite{zou2018beating, chen_-chip_2023, chiuri_hyperentangled_2010, haas_entangled_2014, haffner_scalable_2005, hume_preparation_2009, noguchi_generation_2012, prevedel_experimental_2009, toyoda_generation_2011, wieczorek_experimental_2009, yu_observation_2019} offers the prospect of studying our proposal experimentally in a laboratory.}
 
\textit{Acknowledgements.}
We acknowledge very helpful discussions with Allen Zang and Tian Zhong, and Quinn Langfitt for assistance with the calculations. This material is based upon work supported by the U.S. Department of Energy Office of Science National Quantum Information Science Research Centers. S.K.G. and Z.H.S. were supported by the Q-NEXT Center. Work performed at the Center for Nanoscale Materials, a U.S. Department of Energy Office of Science User Facility, was supported by thee U.S. DOE Office of Basic Energy Sciences, under Contract No. DE-AC02-06CH11357. A.~S.~was supported by QuEST grant No Q-113 of the Department of Science and Technology, Government of India. The results in this work were obtained from simulations using $\sim$1000 CPU-hours on the Bebop computing cluster at Argonne National Laboratory.

\appendix

\section{Numerical Simulation Details}
\label{sec:numerics}

Here we provide additional details about the numerical simulations of a spin-boson system performed for the main text.
The codes for these simulations are available at Ref.~\cite{perlin2023spinboson}.
% MAP: our data is not actually from qutip.  We benchmark against qutip though in small systems to ensure correctness.  I added a sentence to say so below.
%The density matrix is represented and evolved using using the Quantum Toolbox in Python QuTiP~\cite{johansson2012qutip,JOHANSSON20131234} and our own codes that employ QuTiP may be found in Ref. Ref.~\cite{perlin2023spinboson}.

For a system of $N$ spins, we limit the boson excitation number to $N$.
Spin-permutation symmetry then allows us to expand the density matrix in the form \cite{chase2008collective, shammah2018open}
\begin{align}
  \rho = \sum_{J,n,m,\ell,k}
  \rho_{Jnm\ell k} \op{J,n}{J,m} \otimes \op{\ell}{k},
  \label{eq:rho_full}
\end{align}
where $J\in\{\frac{N}{2},\frac{N}{2}-1,\cdots\}$ is a (nonnegative) total spin length, $n,m\in\{J,J-1,\cdots,-J\}$ are spin projections onto the spin quantization axis, and $\ell,k\in\{0,1,\cdots,N\}$ are boson excitation numbers.

The naive numerical representation of $\rho$ in Eq.~\eqref{eq:rho_full} is with a 5-dimensional array, with dimensions $(\lceil N/2 \rceil, \lceil N/2 \rceil, \lceil N/2 \rceil, N+1, N+1)$.
However, about $2/3$ of the elements in this array (namely, the elements with $\abs{n},\abs{m}>J$) are guaranteed to be zero.
In practice, we therefore combine $(J,n,m)$ into a single integer index $i$, and flatten the density matrix into the three-dimensional array
\begin{align}
  \oket\rho = \sum_{i,\ell,k}
  \rho_{i\ell k} \ket{i,\ell,k},
\end{align}
Operators acting $\rho$ are then split into ``left-acting'' and ``right-acting'' variants to infer their action on $\oket\rho$.
The left- and right-acting boson ladder operators, for example, act on $\rho$ as
\begin{align}
  a\rho &\coloneqq a_L \oket{\rho}
  = \sum_{i,\ell,k}
  \rho_{i\ell k} \sqrt{\ell} \ket{i,\ell-1,k},
  \label{eq:a_rho_L}
  \\
  \rho a^\dag &\coloneqq a^\dag_R \oket{\rho}
  = \sum_{i,\ell,k}
  \rho_{i\ell k} \sqrt{k} \ket{i,\ell,k-1},
  \label{eq:a_rho_R}
\end{align}
 which implies that $a_L = \1 \otimes a \otimes \1$, and $a_R^\dag = \1 \otimes \1 \otimes a$, 
 \begin{align}
   a_L = \1 \otimes a \otimes \1,
   &&
   a_R^\dag = \1 \otimes \1 \otimes a,
   \label{eq:a_LR}
 \end{align}
where $\1$ is the identity operator, and by slight abuse of notation the annihilation operators $a$ on the right-hand side of Eq.~\eqref{eq:a_LR} act on the Hilbert space of the boson alone (whereas the operators $a$ on the left-hand side of Eqs.~\eqref{eq:a_rho_L}-\eqref{eq:a_rho_R} act on the joint spin-boson Hilbert space).

Left- and right-acting collective spin operators act on the first tensor factor of $\oket\rho$, and can be constructed from
\begin{align}
  S_z &= \sum_{J,m} m \op{J,m}, \\
  S_\pm &= \sum_{J,m} \sqrt{J(J+1)-m(m\pm1)} \op{J,m\pm1}{J,m},
\end{align}
where the sums implicitly restrict $\abs{m}\le J$, and the left- and right-acting variants of $S_z,S_\pm$ on $\oket\rho$ can be constructed using the same index mapping used to map $\op{J,n}{J,m} \to \ket{i}$ for $\rho\to\oket\rho$.

The final ingredient necessary to simulate the results in the main text is the evaluation of dissipation terms of the form
\begin{align}
  \L[s] = \sum_{j=1}^N \D\sp{s^{(j)}}(\rho),
\end{align}
where $s = s_\1 + s_+ \sigma_+ + s_- \sigma_- + s_z \sigma_z$ 
% \begin{align}
%   s = s_\1 + s_+ \sigma_+ + s_- \sigma_- + s_z \sigma_z,
% \end{align}
is a single-qubit operator defined in terms of the Pauli matrices $\sigma_\alpha$.
The effect of these dissipation terms in the $(J,n,m)$ basis were worked out in Ref.~\cite{chase2008collective}, and specifically in Eqs.~(39)--(46) therein.

In the special case that $s=\sigma_-$, the relevant results in Ref.~\cite{chase2008collective} reduce to
\begin{align}
  \L[s] = \sum_{j=1}^N \sigma_-^{(j)} \rho \sigma_+^{(j)} - \frac14 (S_z \rho + \rho S_z),
\end{align}
where
\begin{align}
  &\sum_{j=1}^N \sigma_-^{(j)} \op{J,n}{J,m} \sigma_+^{(j)}
  \nonumber\\
  &= \frac1{2J} \p{1 + \frac{\alpha_N^{J+1}}{d_N^J} \frac{2J+1}{J+1}}
  A_-^{J\ell} \op{J,\ell-1}{J,k-1} A_-^{Jk}
  \nonumber\\
  &\quad+ \frac{\alpha^J_N}{2J d^J_N} B_-^{J\ell} \op{J-1,\ell-1}{J-1,k-1} B_-^{Jk}
  \nonumber\\
  &\quad+ \frac{\alpha_N^{J+1}}{2(J+1) d_N^J} D_-^{J\ell} \op{J+1,\ell-1}{J+1,k-1} D_-^{Jk},
\end{align}
where
\begin{align}
  \alpha_N^J &= {N \choose N/2 - J}, \\
  d_N^J &= {N \choose N/2 - J} \times \frac{2J+1}{N/2 + J + 1}, \\
  A_-^{J\ell} &= \sqrt{(J+\ell)(J-\ell+1)} \\
  B_-^{J\ell} &= -\sqrt{(J+\ell)(J+\ell-1)} \\
  D_-^{J\ell} &= \sqrt{(J-\ell+1)(J-\ell+2)}.
\end{align}
These ingredients are sufficient to determine the state $\rho$ at all times by numerically integrating the equations of motion in Eq.~\eqref{Lindblad2} of the main text.
We perform this numerical integration using the \texttt{DOP853} method of \texttt{scipy.integrate.solve\_ivp} \cite{scipy}, with relative and absolute error tolerances of $\epsilon_{\mathrm{rtol}} = \epsilon_{\mathrm{atol}} = 10^{-10}$.
We benchmark the correctness and accuracy of our methods against small-scale ($N\le 5$) simulations performed with the Quantum Toolbox in Python QuTiP~\cite{johansson2012qutip, JOHANSSON20131234}.

Denoting the state $\rho$ at time $t$ after evolving under the equations of motion in Eq.~\eqref{Lindblad2} with coupling constant $g$ as $\rho(g,t)$, the QFI of $\rho$ with respect to $g$ at time $t$ is then determined by evaluating Eq.~\eqref{fisher1} using finite differences, namely:
\begin{align}
  F(g,t) = 2 \sum_{a,b}
\frac{\abs{\braket{\lambda_a|\partial_g\rho(g,t)|\lambda_b}}^2}{\lambda_{a}+\lambda_{b}}
  \label{eq:fisher_g_t}
\end{align}
where the sum is taken over values of $a,b$ for which $\lambda_a + \lambda_b\ne0$, and
\begin{align}
  \partial_g \rho(g,t) \approx \frac{\rho(g+\delta g/2, t) - \rho(g-\delta g/2, t)}{\delta g}
\end{align}
with a positive $\delta\ll1$.
We set $\delta = 10^{-3}$ in our simulations.

The structure of the equations of motion, Eq. \ref{Lindblad2},  and Hamiltonian, Eq. \ref{Hamil}, lead to useful scalings.  For simplicity, consider the rotating frame limit that we use in the main text, $\omega_q$ = $\omega_c$ = 0. One can easily see that if one expresses the equations of motion in terms of a dimensionless time, $\tilde{t} = g t$, then 
$d \rho /d\tilde{t}$ involves the dissipator parameters divided by $g$, i.e., $\tilde{\kappa}$ = $\kappa/g$, $\tilde{\gamma}$ = $\gamma/g$ and the qubit-cavity Hamiltonian coupling becomes $\tilde{g}$ = 1. Thus, calculations with $\tilde{g}$ =1 can yield information about results for all $g$ if one reinterprets the dissipator  parameters as being divided by $g$.  It is also easy to show that the QFI for a particular value of $g$, $F(g)$,  can be related to the QFI calculated with respect to $\tilde{g}$ at $\tilde{g}$ = 1, $F(\tilde{g})|_{\tilde{g}=1}$ by:  
$F(g) \times g^2$ = 
$F(\tilde{g})|_{\tilde{g}=1}$. 

A final subtlety of our simulations is the treatment of floating-point errors, which can result in numerical instabilities of $F(g,t)$ due to the denominator $\lambda_a+\lambda_b$ in Eq.~\eqref{eq:fisher_g_t}.
Specifically, floating-point errors cause negative eigenvalues of $\rho$ to pick up a small nonzero value.
To mitigate instabilities from small nonzero eigenvalues in the denominator of Eq.~\eqref{eq:fisher_g_t}, we identify $\epsilon_{\mathrm{err}} = \abs{\min_a \lambda_a}$ as the scale of floating-point errors in $\rho$.
We note that $\min_a \lambda_a$ is always negative in practice, since floating-point errors cause each zero eigenvalue of $\rho$ to become negative with $\sim50\%$ probability.
When evaluating $F(g,t)$, we then set eigenvalues less than $10\epsilon_{\mathrm{err}}$ to $0$, and sum over all values of $a,b$ for which $\lambda_a+\lambda_b > 10^{-8}$.

\section{Time-Dependence of QFI for Varying System Size and Probe States}
\label{sec:timedependence}

\begin{figure}
    \centering
    \includegraphics[width=0.9\linewidth]{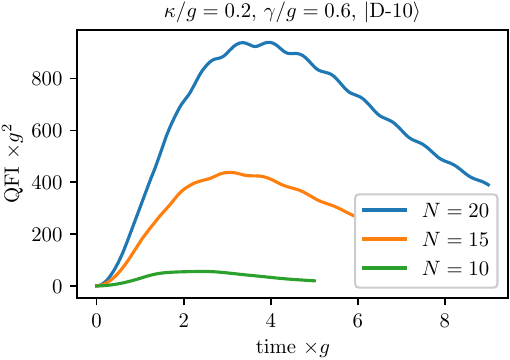}
    \includegraphics[width=0.9\linewidth]{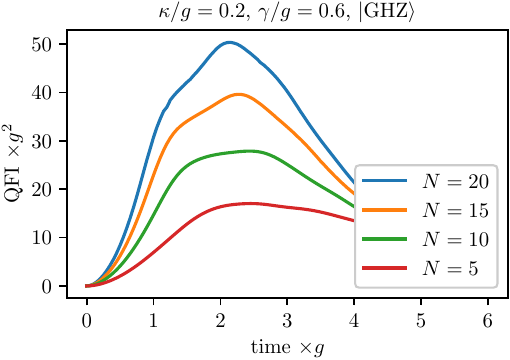}
    \includegraphics[width=0.9\linewidth]{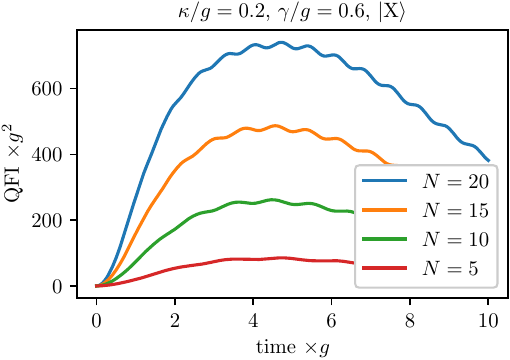}
    \caption{Time dependence of QFI for different probe states. Each of the figures shows how the maximum value of the QFI changes with time and also with varying $N$.}
    \label{timedependence}
\end{figure}

Here we provide a few more examples of how the optimal time for performing measurements on the quantum system changes as we change the number of qubits and the probe states. It can be clearly seen in Fig \ref{timedependence} that for different initial states the optimal time for performing the measurement changes as we change the number of qubits. 

\section{Computations with Time as a Resource}
\label{sec:QFIovertime}

\begin{figure}
    \includegraphics[width=7.5cm, height=5.5cm]{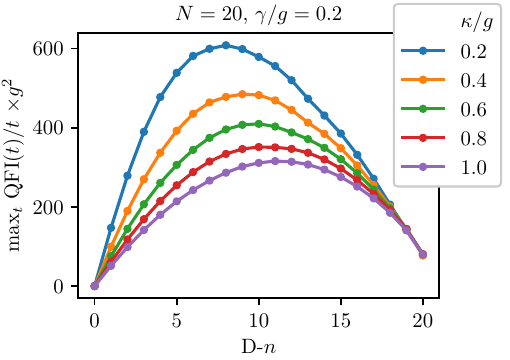}
    \caption{Variation of the maximum QFI$(t)/t$ (with respect to time) with Dicke excitation number for $N$ = 20 qubits and varying resonator decay rate, $\kappa$, for fixed qubit decay rate, $\gamma$.}
    \label{tMaxQFIvsn}
\end{figure}

\begin{figure}
  \includegraphics[width=7.2cm, height=5.3cm]{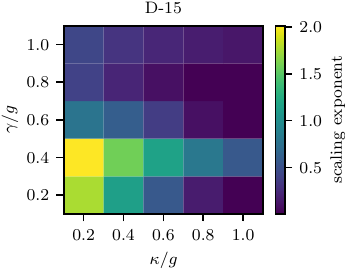}
  \includegraphics[width=7.2cm, height=5.3cm]{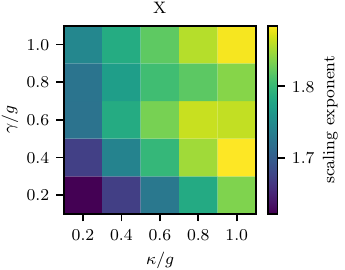}
 \caption{Scaling exponents $b$ from a three-parameter fit to $\max_t\mathrm{QFI}(t,N)/t\sim a N^b + c$, where $N$ is the qubit number, for the Dicke-15 state and the polarized-X state.}
  \label{scalingqfitime}
\end{figure}

Let $T$ be the total time the expermienter has and the $t$ be the measurement time. Then the total number of applications of the quantum probe is $M = T/t$. The Cramer-Rao bound can now be rephrased as,
\begin{align}
  \delta \hat \theta ^2 \geq \frac{1}{T F(\theta)/t}.
\end{align}
The quantity that we optimize with respect to time then becomes the $F(t,\theta)/t$ instead of just $F$. Our results are provided in the Figs \ref{tMaxQFIvsn} and  \ref{scalingqfitime}. 

\textcolor{black}{In Fig.~\ref{tMaxQFIvsn}, the variation of QFI$(t)/t$ with excitation number for the Dicke-20 state for different noise parameters are shown. We see that the behaviour of QFI$(t)/t$ mirrors that of QFI in Fig.~\ref{MaxQFIvsn}, indicating that our conclusions regarding the optimal excitation number for Dicke states holds whether we consider time as a resource or not. The same is true for the scaling of the QFI at optimal times as a function of $N$ as well, in Fig.~\ref{scalingqfitime}. Our main conclusions therefore remain unchanged if we treat the total time $T$ as a resource.}

\pagebreak
\bibliography{references}% Produces the bibliography via BibTeX.
\pagebreak

The submitted manuscript has been created by UChicago Argonne, LLC, Operator of Argonne National Laboratory (``Argonne''). Argonne, a U.S. Department of Energy Office of Science laboratory, is operated under Contract No. DE-AC02-06CH11357. The U.S. Government retains for itself, and others acting on its behalf, a paid-up nonexclusive, irrevocable worldwide license in said article to reproduce, prepare derivative works, distribute copies to the public, and perform publicly and display publicly, by or on behalf of the Government. The Department of Energy will provide public access to these results of federally sponsored research in accordance with the DOE Public Access Plan (\url{http://energy.gov/downloads/doe-public-access-plan}).
\end{document}